\documentstyle[12pt]{article}
\newcommand{\rf}[1]{(\ref{#1})}
\newcommand{\beq}{\begin{equation}}
\newcommand{\eeq}{\end{equation}}
\newcommand{\g}{\gamma}
\renewcommand{\l}{\lambda}
\renewcommand{\b}{\beta}
\renewcommand{\a}{\alpha}
\newcommand{\n}{\nu}
\newcommand{\m}{\mu}
\renewcommand{\d}{\delta}

\newcommand{\ep}{\varepsilon}

\newcommand{\sg}{\sigma}
\newcommand{\bea}{\begin{eqnarray}}
\newcommand{\eea}{\end{eqnarray}}
\renewcommand{\ni}{\noindent}
\newcommand{\oh}{\frac{1}{2}}

\newcommand{\CT}{\mbox{${\cal T}$}}

\newcommand{\CF}{\mbox{${\cal F}$}}
\newcommand{\CG}{\mbox{${\cal G}$}}

\newcommand{\no}{\nonumber}

\newcommand{\del}{\partial}
\newcommand{\gsim}{\stackrel{\textstyle >}{\sim}}
\catcode`@=11
\def\addtoreset#1[#2]{\@addtoreset{#1}{#2}}
\addtoreset{equation}[section]

\begin{document}
\topmargin -20mm
\oddsidemargin 5mm
\headheight 0pt
\topskip 0mm

\addtolength{\baselineskip}{0.20\baselineskip}
\hfill NBI-HE-92-40

\hfill June 1992
\begin{center}

\vspace{12pt}
{\large \bf
THE THEORY OF DYNAMICAL RANDOM SURFACES WITH EXTRINSIC CURVATURE}
\end{center}

\vspace{12pt}

\begin{center}
{\sl J. Ambj\o rn }

\vspace{12pt}

The Niels Bohr Institute\\
Blegdamsvej 17, DK-2100 Copenhagen \O , Denmark\\

\vspace{12pt}

{\sl A. Irb\"{a}ck}

\vspace{12pt}

Theory Division, CERN - CH 1211 Geneva 23, Switzerland\\

\vspace{12pt}

{\sl J. Jurkiewicz}{\footnote{Partially supported by the KBN grant
no. 2 0053 91 01}}

\vspace{12pt}

Institute of Physics, Jagellonian University,\\
 ul. Reymonta 4, PL-30 059 Krak\'{o}w~16, Poland

\vspace{12pt}

{\sl B. Petersson}

\vspace{12pt}

Fakult\"{a}t f\"{u}r Physik, Universit\"{a}t Bielefeld,
D-4800 Bielefeld 1, Germany\\

\end{center}

\vfill

\begin{center}
{\bf Abstract}
\end{center}

\vspace{12pt}

\ni
We analyze numerically the critical properties of a two-dimensional
discretized random surface with extrinsic curvature
embedded in a three-dimensional space.
The use of the toroidal topology enables
us to enforce the non-zero external extension
without the necessity of defining a boundary and allows us to
measure directly the string tension. We show that
a phase transition from the crumpled phase
to the smooth phase observed earlier for a spherical topology appears
also for a toroidal surface
for the same finite value of the coupling constant of the
extrinsic curvature term.
The phase transition is characterized by the vanishing
of the string tension. We discuss the possible non-trivial continuum
limit of the theory, when approaching the critical point.
Numerically
we find a value of the critical exponent $\n$ to be between .38 and .42.
The specific heat, related to the extrinsic curvature term seems not
to diverge (or diverge slower than logarithmically) at the critical point.

\vspace{24pt}
\vfill

\newpage

\section{Introduction}

The theory of random walks has provided us with a powerful
link between statistical mechanics and euclidean field theory.
An euclidian field theory can be expanded in a series of intersecting
random walks and a number of rigorous inequalities can be proven using
the random walk representation. In addition various aspects of universality
in regularized field theory can easily be understood from the corresponding
universality of the random walks. It has further  been possible to develop
a theory of  random walks, which allows
the description of fermions  in a geometrical way.

The theory of random surfaces ought to provide us with even stronger
tools in the study of string theories. String theory in its first
quantized version is nothing but the theory of a specific kind of
{\it free} random surfaces. This has been substantiated during
the last couple of years, but many questions remain unanswered.
One of the greatest puzzles is that the formulas derived by
Knizhnik, Polyakov and Zamolodchikov \cite{kpz} using conformal field theory
seem to make no sense for strings embedded in "physical" dimensions
$(d >1)$. In more general terms we do not know how to couple two-dimensional
gravity to conformal field theories with central charge $c >1$.
On the other hand it seems to be no problem to formulate statistical
theories of random surfaces in physical dimensions. The use of these
random surface theories in the context of strings is hampered by our
lack of understanding of scaling and universality properties of the random
surface theories themselves. The situation here is quite different
from the random walk case. The present work is an attempt to
clarify the situation for a particular class of random surface models.
Let us briefly motivate why this class is interesting.

The class of random surface theories we have in mind is one of freely
intersecting surfaces with an action having an area term
and an extrinsic curvature term. Such a model can be viewed as a
simplified model of physical membranes. The statistical aspect enters
due to thermal fluctuations and the statistical fluctuations may drive
the system to a point where the effective surface tension vanishes and
the membranes are dominated by their curvature energy rather than their
surface tension. This is the case for fluid/fluid interfaces and also
for the so-called amphiphilic membranes, which are membranes formed
when amphiphilic molecules are brought into contact with water and
form bilayers by orienting their polar parts towards water and their
oily hydrophobic tails away from the water. Of course such surfaces
are not allowed to self-intersect and from this point of view our
models can only be viewed as toy models for real membranes. However,
since the seminal work of Helfrich \cite{helfrich}
such toy models have received a
lot of attention \cite{klein}.

The random surface models with extrinsic curvature are also interesting
as effective theories of strings. In this context they were first suggested
by  Polyakov \cite{pol1}. He had in mind an effective string theory which
could be equivalent (at least at long distances) to QCD with heavy
quarks. However, it is now clear that fermionic string theories can
give rise to effective bosonic string theories which have extrinsic
curvature terms, and in this context the surfaces {\it should} be allowed
to self-intersect. In the case of the superstring it is possible to
integrate out the world-sheet fermions \cite{wiegmann}.
After this integration two types of terms are
produced, which both depend on the {\it extrinsic } geometry of the
world sheet~:
\beq
S_{eff}= S_{bosonic}+  W_k (A^{(n)})+
 \frac{\tau}{8} \int d^2\xi\sqrt{g_{ind}}\;
\left\{ (e^\m_\a \partial_\b e^\m_\g)^2 + (D_\a n^\m_i)^2 \right\}
\label{1}
\eeq
Here $g_{ind}$ refers to the metric induced by target space,
$n^\m_i\;,\;i=1,\ldots,d-2$ are normals to the surface, $e^\m_\a\;
\; \a=1,2$ are the tangents and $D_\a$ denotes the covariant derivative
with respect to the connection $A^{(n)}$ in the normal bundle~:
\beq
A^{(n) \a}_{ij}=n^\m_i\partial_\a n^\m_j - n^\m_j \partial_\a n^\m_i.
\label{2}
\eeq
$\tau$ is a Dynkin factor coming from the fermionic representation and
finally $W_k (A^{(n)})$ denotes the Wess-Zumino action~:
\beq
W_k (A^{(n)}) = \frac{ik}{8} {\rm Tr}\;\left( \frac{1}{2}
\int d^2 \xi \; A \wedge A +\right.
 \left. \frac{1}{3\pi} \int_D d^3 x \;
A \wedge A \wedge A \right)
\label{3}
\eeq
where we have used the notation $A^\a = A^{(n) \a}_{ij} M_{ij}$, $M_{ij}$
being the generators of $ {\rm SO} (d-2)$ and $D$ a three-dimensional
disc bounded by the world sheet.

One class of terms depends very explicitly on the extrinsic geometry
of the world sheet and is minimized by smooth, flat surfaces.
If $K(\xi)$ denotes the extrinsic curvature of the world sheet we have
\beq
K(\xi) = \frac{1}{r_1 (\xi)} + \frac{1}{r_2 (\xi)}    \label{4}
\eeq
where $r_1 (\xi)$ and $r_2(\xi)$ are the principal curvatures of the
surface. Since
\beq
\int d^2 \xi \sqrt{g_{ind}}\; (D_\a n_i)^2 =
\int d^2 \xi \sqrt{g_{ind}}\; K^2 (\xi) \label{5}
\eeq
this term clearly favours smooth surfaces. The same is the case
for the term involving the tangents.

As for the Wess-Zumino-like term $W(A)$, the effect is not so clear.
In euclidean space-time the term is purely imaginary.
It is interesting to compare with the situation in the case of random walks.
What we have in mind is a ``supersymmetric'' random walk, where we
have introduced world-line fermions. In this case it is again possible
to integrate out the fermions. The result is a Wess-Zumino-like term on the
world-line and it is the analogue of the term $W_k(A)$ defined in eq. \rf{3}.
In the random walk case the effect of the Wess-Zumino term can be analyzed in
detail (see \cite{book} for a continuum treatment and \cite{adj} for a
random walk approach). Heuristically the result can be described as follows:
The amplitudes of back-tracking random walks tend to cancel due to the
phase factor coming from the Wess-Zumino term and we are effectively
left with a class of random walks which are much smoother than the
generic random walks of a scalar particle. In effect the short distance
Hausdorff dimension of the random walk is reduced from two to one, and
the corresponding short distance behaviour of the corresponding
propagator changes from being like
$1/k^2$ to $1/k$. As first noticed by Polyakov and proven in the context
of random walks in \cite{adj}, the scaling limit of this fermionic
random walk leads to a representation of the massive Dirac propagator.
In this case the effect of world-line supersymmetry and  the corresponding
Wess-Zumino term is clearly to favour ``smooth'' curves. It is natural
to conjecture that the two-dimensional Wess-Zumino term \rf{3} has a similar
effect.

As we see all  additional terms in \rf{1} act in favour of
smoother surfaces as compared to the class of surfaces singled
out by the standard bosonic term $S_{bosonic}$. This opens up
the possibility of an intuitive and simple understanding of the
tachyon problem of the bosonic string and its cure in the case
of the superstring. The appearance of tachyons in a bosonic random
surface theory is somewhat different from that  based on formal
continuum manipulations. By definition there can be no tachyons
in a theory of random surfaces where each surface is given a positive
weight and which satisfy the principle of reflection positivity.
However, the scaling limit of such discretized, regularized theories
might be pathological from the point of view of string theory. This
is precisely what happens for the ordinary bosonic string theory
where the dimension $d$ of target space is larger than one.
One can prove that the string tension does not scale \cite{ad} for such
theories and as a consequence the scaling limit is not really that
of a ``surface'' theory, but rather that of a theory of so-called
branched polymers consisting of a minimal surface (depending on the
boundary conditions) from which the only allowed fluctuations are thin
``branches'' which carry no area. Maybe somewhat contrary to intuition
the entropy of such surfaces is large compared to that of ``smooth''
surfaces and completely determines the scaling properties of the
``surface'' theory. An obvious cure is to put in by hand additional
terms in the action which suppress the ``spikes'' and
favour a smoother class of surfaces and
this is precisely what is done by imposing world-sheet supersymmetry
and integrating out the world-sheet fermions as indicated in \rf{1}.

We know there are no tachyons in the superstring theories. However, we
also know that  in the usual continuum description the absence is due to
a delicate cancellation between bosonic and fermionic degrees
of freedom. From this point of view it is intriguing how a discretization
of the action \rf{1} and a subsequent scaling limit is able to capture
this. The obvious answer would be: by universality. But although it
makes no sense to derive an effective action like \rf{1}, except in the
special dimensions $d=3,4,6$ and 10, where a classical superstring
theory can be defined, the effective action itself makes sense in any
dimension and one can consider the corresponding discretized theory  and
possible scaling limits. Appealing to analytic results there should
be a difference between the theories, depending on the dimension of
target space. At the moment we have no understanding of the physics
leading to  such a difference. The concept of universality triggers
a related question: How far are we able to modify a discretized version
of the effective action \rf{1} and still stay in the same universality
class. Due to the imaginary Wess-Zumino term we are clearly discussing
a rather unconventional class of theories from the point of view
of statistical mechanics.  Nevertheless it does not necessarily mean
that one cannot apply the conventional machinery of statistical mechanics.
The above mentioned example of the fermionic random walk provides
documentation for this and shows that one can indeed talk about
universality classes \cite{adj}. It would be very convenient if we were
able to drop completely the imaginary Wess-Zumino term and still stay
in the same universality class. At a first glance it looks unlikely, again
appealing to the fermionic random walk example. One can analyse the random
walk where one has both an extrinsic curvature term like \rf{5} and the
Wess-Zumino term. The result is as follows: if we choose the
coefficient of the Wess-Zumino term to be zero we will in general be in the
universality class of the ordinary random walk (and the corresponding
scaling limit will be that of a free scalar particle)\footnote{The only
exception is if we take a scaling limit where we scale the bare coupling
constant of the  extrinsic curvature term to infinity. It is possible
to do this in such a way that we get what can be called a rigid
random walk\cite{adj1}.}. As soon as we take the coefficient of the
Wess-Zumino term to be non-zero we are driven to the fermionic random walk,
where the scaling limit  leads to a massive Dirac propagator.
The same could be
true for the effective action \rf{1}. However, since the world-sheet
is two-dimensional while the world-line is one-dimensional we are discussing
vastly different theories from the point of view of statistical
mechanics. The structure of phase transitions is much richer in two
dimensions, and although a one loop calculation seems to support the
conclusion reached in the random walk case \cite{pol}, this is a purely
perturbative argument, based on assumptions which will not be satisfied
in case we have a non-perturbative phase transition. A priori it is possible
that one could drop the Wess-Zumino term and still stay in the same
universality class.

The above discussion is closely linked to another  discussion of
possible critical behaviour of membranes (or random surfaces) with
extrinsic curvature terms. It is believed that one {\it has} a transition
from so-called ``crumpled'' surfaces to smooth surfaces in the case
of crystalline surfaces. Crystalline surfaces can be viewed as
membranes where the individual molecules have a fixed connectivity, i.e.
their neighbours are fixed, contrary to the situation for fluid membranes.
If the coefficient of the extrinsic curvature term is zero such
surfaces will be ``crumpled'' (the Hausdorff dimension for the
statistical ensemble of surfaces is infinite),
at least in the idealized situation where
they are allowed to intersect freely. However, for a finite value of the
extrinsic curvature coupling there seems to be a phase transition, and
for couplings above this critical value the surfaces  are smooth
(the Hausdorff dimension for the statistical ensemble of surfaces is two).
No rigorous proof of this phase transition exists, but numerous
numerical results seem to confirm the existence of such a transition,
and it seems to be a second order transition for the kind of extrinsic
curvature term we are going to use in this work. In the beginning some
confusion surrounded this transition. Depending on the details of the
discretized version of the extrinsic curvature term used, the transition
was classified as first, second or third order, respectively. It is now
understood that only the discretization which seems to lead to a second
order transition is not ``pathological''. The other actions used, although
formally equivalent for smooth surfaces, led to singular surface
configurations if the surface was allowed to fluctuate wildly, as in fact
happened in the ``crumpled phase''. This complicated situation highlights
the possibility of a nontrivial phase structure in random surface theories
and the care one has to exercise in order to choose a correct discretized
action. One can now imagine a situation where the attachment of the
molecules to specific neighbour molecules in the crystalline surface
is gradually decreased. The crystalline structure is allowed to ``melt''.
Ultimately one will end up with a fluid membrane where the individual
molecules can move freely. Will the second order phase transition extend
all the way to a fluid membrane or can one trust the one loop calculation
\cite{pol,helfrich} done for the fluid membrane case,
which indicated that there is no transition\footnote{We thank John Wheater
and Fran\c{c}ois David, who independently
suggested this interpolating scenario to us.}~? Computer simulations
first started by Catterall and later repeated by other groups point
to the existence of a ``crumpling'' transition even in the
case of fluid membranes, but contrary to the situation for crystalline
surfaces numerical simulations and analytic arguments contradict
each other in the fluid membrane case.

We hope to have convinced the reader that the critical properties of
random surfaces are a fascinating topic of importance in vastly different
areas of physics, but that it is an area
where only little is known at present.
The aim of the present article is to develop some theoretical concepts,
adequate for the description of the critical phenomena of random surfaces,
and by extensive numerical simulations try to answer some of the
questions raised above.

The paper is organized as follows: In section 2 we discuss the model and
define the observables, which we use in the canonical numerical simulations.
We also relate them to the standard grand canonical definitions of the
mass gap and string tension. Scaling properties of these observables are
discussed in section 3. In section 4 we describe the system with twisted
boundary conditions, which was used in all our numerical work to measure
the string tension and the mass gap. Numerical results are presented in
section 5, which is divided into four sub-sections, concerning the
specific heat, the mass gap, the string tension and
the radius of gyration. We conclude the paper with a discussion
in section 6.

\section{The action and observables}

Let us in this section define the action which we are going to
use and discuss the observables which will allow us to investigate
the critical properties of the system. We want to approximate the
continuum surfaces discussed in the introduction with piecewise
flat surfaces. The intrinsic parameter space is then identified
with an abstract triangulation of appropriate topology, defined entirely
by its vertices $i$ and links $<ij>$. The embedding in a $D$-dimensional
euclidean target space is a map assigning
to each link $<ij>$ a vector $X_{ij}^{\m},\m = 1...D$ in the target
space. The numerical results presented in this paper are related solely
to the case $D =3$.
Three vectors forming a triangle $i,j,k$ should satisfy
\beq
X_{ij}^{\m} + X_{jk}^{\m} + X_{ki}^{\m} = 0. \label{zero}
\eeq
If the topology is spherical, \rf{zero}
would mean that
\beq
X_{ij}^{\m} = X_j^{\m} - X_i^{\m}, \label{def}
\eeq
where $X_i^{\m}$ denotes coordinates of a vertex $i$ in the target space
and a sum of vectors $X_{ij}^{\m}$ along any closed path would be zero.
For surfaces with handles the embedding may be non-trivial and we will
discuss this further in section 4. In the rest of this section
we will assume that the topology is trivial.

A configuration of the surface is defined by specifying
a triangulation (links
joining the vertices) and by assigning values to vectors
$X_{ij}^{\m}$ satisfying \rf{zero}.
For each
configuration we define the action $S =\b S_G + \l S_C$. $S_G$ is the
gaussian part of the action
\beq
S_G = \oh \sum_{<ij>} (X_{ij}^{\m})^2  \label{gauss}
\eeq
and $S_C$ is the extrinsic curvature part
\beq
S_C = \sum_{<ij>} (1 -  \cos \theta_{ij} ),  \label{extr}
\eeq
$\theta_{ij}$ denoting the angle between the normals to two oriented
triangles $<ijk>$ and $<jil>$ with
a common link $<ij>$. The coupling constant $\b$ of the gaussian part
of the action plays the role of the unit length.

This action is intended to serve as a discretized version of the terms
\beq
\int\; d^2 \xi \sqrt{g} \partial_\a X^\m \partial^\a X^\m+
\l \int d^2 \xi \sqrt{g_{ind}} \; K^2(\xi)  \label{2.1}
\eeq
in \rf{1}, where $g_{ind}$ refers to the metric induced by target space.
Strictly speaking  $S_C$ given by \rf{extr} is not a direct translation of
the last term in \rf{2.1}, since we are not using the induced metric.
However, we do not expect there will be any significant difference (i.e.
the action \rf{2.1} should belong to the correct  universality class),
since both the order of vertices and the shape of triangles in target space
are smooth functions.

Two statistical ensembles have our interest: the canonical ensemble where
the number triangles is kept fixed, and the grand canonical ensemble
where the number of triangles is allowed to fluctuate.

The canonical partition function for a closed surface of
trivial topology is given by
\beq
Z_N(\b ,\l )  =
\sum_{T  \in \CT}\int \prod_{i=1}^{N} dX_i^{\m} \d (\sum X_j^{\m})
   \exp (-S),
\label{aa2}
\eeq
where the first sum is over all possible triangulations of the surface
with $N$ vertices and the delta function is necessary because of the
translational invariance of the action.
We can always scale out $\b$ since $S_C$ is scale invariant and we are
left only with one coupling constant $\l$.
The partition function defines Helmholtz free energy
\bea
\CF (N,\b ,\l )= -\log Z_N(\b ,\l). \label{2.1a}
\eea
Using again the scale invariance of  $S_C$  we have the following
transformation under scaling $X^\m_i \to y {X'}^\m_i$:
\beq
S(X^\m_i ) =y^2 \b S_G({X'}_i^\m)+ \l S_C({X'}_i^\m)  \label{2.2}
\eeq
and Helmholtz free energy will scale as
\beq
\CF (N,\b,\l)=\CF (N,y^2\b,\l)-(N-1)D \log y.  \label{2.3}
\eeq

The grand canonical ensemble will be defined by
\beq
Z(\m,\b,\l) =\sum_N^\infty e^{-\m N} \; Z_N(\b,\l)  \label{2.3a}
\eeq
where $\m$ is a chemical potential for the number of vertices $N$
(or equivalently the number of triangles $N_2$). $\b$ was not a
 dynamical coupling constant for the canonical ensemble. The
same is true in the grand canonical ensemble where it is clear
from   \rf{2.2}-\rf{2.3a} that
\beq
Z(\m,\b,\l)=Z(\m-D\log \b, 1, \l)  \label{2.4}
\eeq
It is now obvious that we can always take $\b=1$, and we will do that
in the following except when explicitly stated differently. The
Gibbs free energy can now be defined by
\beq
\CG (\m,\l) = -\log Z(\m,\l)  \label{2.4a}
\eeq
where we have taken $\b=1$ as announced.

In the case of the the canonical ensemble we reach the thermodynamic
limit by taking the size of the system $N \to \infty$.
This thermodynamic limit will in general depend on the coupling constant
$\l$ and for certain values of $\l$ there might be phase transitions
which, in case they are second order, might serve as points where
we can define a continuum field theory.

In the case of a grand canonical ensemble we will have a critical line
$\m = \m_c(\l)$ in the $(\l,\m)$ coupling constant plane, such that the
theory is defined for $\m > \m_c$. The thermodynamic  limit is
obtained when number of vertices $ N $ diverges, and
$\langle N \rangle \to \infty$ corresponds in the language
of the grand canonical ensemble
to moving close to the curve $\m = \m_c(\l)$ along the $\m$-axis, starting
at large $\m$. The possibility of phase transitions for certain values
of $\l$, as described above in the language of the canonical ensemble
can be addressed in the grand canonical ensemble too.
In this ensemble such a transition
will manifest itself as  points {\it on} the critical line, which separate two
different types of critical behaviour when we move along the critical
line $\m =\m_c(\l)$.

For numerical purposes it is much more convenient to work with a
canonical ensemble. Certain observables are however naturally defined
in the grand canonical ensemble, and in the following we will show
how it is possible to extract information about them using only
the canonical ensemble.

The simplest  observable which gives us information about the nature
of the phase transition is the specific heat with respect to $\l$. It
can be defined directly in the canonical ensemble:
\beq
C(\l) \equiv \frac{\l^2}{N} \frac{ \partial^2 \CF (N,\l)}{\partial \l^2}=
\frac{\l^2}{N} \left( \langle S_C^2 \rangle - \langle S_C \rangle^2 \right)
\label{2.5}
\eeq
In case there is a second order transition at a finite $\l_c$
we should see a singularity of $C(\l)$ as $N \to \infty$ as will be discussed
later.

Even if we observe a second order transition at a finite $\l_c$, it does
not ensure that this transition has anything to do with a string theory.
A minimal  requirement is that the string tension and possible
mass excitations scale to zero when $\l \to \l_c$. The following discussion
is stimulated by the work of David and Leibler \cite{david}.
In order to define the string tension we introduce the grand canonical
ensemble of open surfaces where the boundary is kept fixed.
We imagine the surface will enclose a large area $A$ and denote the
corresponding partition function $Z(\m,\l;A)$.
The partition function will behave as
\beq
Z(\m,\l;A) \sim \exp(-\sg (\m,\l) A),~~~~{\rm for}~~~A \to \infty
\label{2.6}
\eeq
and we can define the string tension as
\beq
\sg (\m,\l) = \lim_{A \to \infty} \frac{\CG(\m,\l ; A) }{A}. \label{2.7}
\eeq

In the same way we can define the mass gap in the theory as the exponential
decay of the grand canonical partition function of surfaces with a boundary
consisting  of two points separated by a distance $L$
\beq
Z(\m,\l; L) \sim \exp(-m(\m,\l) L),~~~~~{\rm for}~~~L \to \infty
\label{2.8}
\eeq
i.e.
\beq
m(\m,\l)= \lim_{L \to \infty} \frac{ \CG (\m,\l; L)}{L}. \label{2.9}
\eeq

In the thermodynamic limit (where $\m(\l) \approx \m_c(\l)$
we have (for any boundary conditions``$B$'') the usual relation between
Helmholtz free energy $\CF$ and the Gibbs free energy $\CG$
\beq
\CG(\m,\l;\mbox{``$B$''}) = \m N + \CF (N,\l;\mbox{``$B$''})
\label{2.10}
\eeq
\beq
N = \frac{\partial \CG(\m,\l ;\mbox{``$B$''})}{\partial \m}  \label{2.11}
\eeq
If we apply the formula to the case where the boundary condition ``$B$''
is the one which is used in the definition of the string tension
we get
\beq
\sg(\m,\l) A = \m N + \CF (N,\l; A)  \label{2.12}
\eeq
from which we conclude that
\beq
\sg(\m,\l) = \frac{\partial \CF(N,\l ;A)}{\partial A} =
\sg(\l,N,A) \label{2.13}
\eeq
with $N$ expressed in terms $\m$ and $A$ by \rf{2.11}
\beq
N = \frac{\partial \sg(\m,\l)}{\partial \m} \, A.  \label{2.14}
\eeq
In the same way we get where the boundary condition ``$B$'' is the one
used in the definition of the mass gap
\beq
m(\m,\l)L = \m N + \CF(N,\l;L) \label{2.12a}
\eeq
from which we conclude that
\beq
m(\m,\l) = \frac{\partial \CF(N,\l; L)}{\partial L} =
m(\l,N,L),  \label{2.15}
\eeq
with $N$ expressed in terms of $\m$ and $L$ by \rf{2.11}:
\beq
N = \frac{\partial m(\m,\l)}{\partial \m}L. \label{2.14a}
\eeq

{\it The advantage of \rf{2.13} and \rf{2.15} is that they allow us to
measure the string tension and the mass gap using only canonical
ensembles}. Further, with the action in question it follows by a
simple scaling argument using \rf{2.2} that
\bea
\frac{\partial \CF (N,\l; A)}{\partial  A} &=&
\frac{\langle S_G \rangle -\oh D (N-1)}{A}  \\ \label{2.16}
\frac{\partial \CF (N,\l; L)}{\partial  L} &=&
\frac{2\langle S_G \rangle - D (N-1)}{L}.   \label{2.17}
\eea
{\it In this way we see that we can measure string tension and mass gap
by a simple measurement of the local observable $S_G$}. To study the
critical behaviour it is convenient to consider an analogue
of the specific heat \rf{2.5}
\beq
\rho(\l,N; A) = \frac{A^2}{N} \;
\frac{\del^2 \CF (N,\l;A)}{\del A^2} =
-\frac{\langle S^2_G \rangle -\langle S_G \rangle^2- \oh D(N-1)}{N}
\label{2.18}
\eeq

\vspace{12pt}

\noindent
In deriving \rf{2.16} and \rf{2.18} we have used only scaling arguments,
so the string tension should not depend on the shape of the fixed boundary
loop. This will only be true when the two linear dimensions of the
enclosed area are comparable. For elongated loops the behaviour will
eventually change, since it will be difficult to distinguish the physics
associated with the area from the (different) physics associated with
the perimeter.

\section{Scaling properties of the observables}

The aim of the numerical study  is to find and understand the
critical properties of the system. For a canonical system with a fixed  number
of points $N \to \infty$ the system is believed to undergo a phase
transition for a finite value of the coupling constant $\l$. For the target
space dimension $D=3$ this phase transition seems to be of the second order.
A standard method to localize the critical point is to study the {\it specific
heat} $C(\l )$ \rf{2.5}.
For a second order phase transition the function $C(\l)$ has a singularity at
$\l = \l_c$ in the thermodynamic limit $N \to \infty$. For finite values of
$N$,
$C^N(\l)$ is a continuous function of $\l$, which develops a maximum at some
$\l = \l_c^N$. Let the maximum of $C^N(\l)$ be
\beq
C_{max}^N = C^N(\l_c^N).
\eeq
{}From the standard finite-size scaling arguments ({\it e.g.}\cite{fsscal}) we
expect for $N \to \infty$ that $\l_c^N \to \l_c$ with
\bea
|\l_c^N - \l_c| &\sim& \left(\frac{1}{N}\right)^{\ep}, \\ \no
C_{max}^N &=& A + B N^{\omega},
\eea
where $\ep$ and $\omega$ are some critical exponents.
The increase of $C_{max}^N$
with growing $N$ is considered to be a signal of the second order phase
transition unless $\omega=1$, in which case the transition
{\it could} be first order. We will measure $C^N(\l)$ by means of
\rf{2.5}. However, it is important to stress, as is apparent from
\rf{2.5}, that $C(\l)$ is not the integral of a simple normal-normal
correlation function and consequently there is no a priori reason
that $C(\l)$ should diverge even if the normal-normal correlation length
diverges and in fact our numerical results are compatible with no or less than
a logarithmic growth of the specific heat with $N$.

For the string tension measurement we can use the canonical
ensemble of open surfaces with a fixed boundary which we will
usually take to be a rectangular $L_1 \times L_2$ loop enclosing a given
area $A$, which is not to be confused with the area of the surface.
As explained in the next section we actually implement the
boundary conditions differently, but for the discussion of scaling
properties the above definition will be sufficient.

The string tension is known not to scale to zero for $\l =0$ and
$\m \to \m_c(\l=0)$ \cite{ad}. The thermodynamic limit in the
grand canonical ensemble is obtained when
\beq
\m_R\equiv \m-\m_c(\l) \to 0  \label{3.1}
\eeq
and it is therefore natural to expect a general behaviour
\beq
{\sg}(\l,\m) = \sg_0(\l) + d(\l)\m_R^{2\n(\l)}
\label{dnew4}
\eeq
with some critical exponent $\n(\l)$. From \rf{2.14}, which tells us
that $N/A = \partial \sg(\m,\l)/\partial \m$
we then deduce
\beq
N \sim \m_R^{2\n(\l) -1} A.
\label{dnew5}
\eeq
It implies that the quantity $r = A/N$ can be expressed in terms of $\m_R$
and using eq. \rf{2.13} we see that the string tension $\sg(\l,N,A)$,
defined in the canonical ensemble, in the critical region for a fixed $\l$
should depend on $N$ and $A$ only through $r$:
\beq
\sg (\l,N,A) = \sg(\l,r)
= \sg_0(\l) + d(\l)r^{2\n(\l)/(1 - 2\n(\l))}. \label{dnew6}
\eeq
In the derivation of \rf{dnew6} we assumed of course
that the finite-size effects
can be neglected. Formula \rf{dnew6} will be used to extract from the numerical
simulations the critical
behaviour of $\sg_0(\l)$ and the value of the critical index $\n(\l)$. The
continuum
limit is related to the small-$r$ behaviour of \rf{dnew6}, where the
finite-size  corrections may be important. This problem will be discussed in
the section 5.

Numerical results on $\sg(\l,r)$ can be used to get an
estimate of the critical exponent $\n$, but the behaviour of ${\sg}(\l,\m)$
given by \rf{dnew4} does not guarantee that
the continuum theory will have a finite physical string tension.
As the thermodynamic limit for a fixed $\l$ is approached for
$\m_R \to 0$ we expect that the possible scaling behaviour is extracted
from
\beq
\sg(\l,\m) A = \sg_{phys} (\l) A_{phys}  \label{aux1}
\eeq
where the {\it physical string tension} $\sg_{phys}(\l)$ and the
{\it physical area} $A_{phys}$ are kept fixed for $\m_R \to 0$.
We further want the ``bare'' area $A$ to go to infinity in order
not to deal with lattice artifacts. These requirements clearly demand
that $\sg_0 (\l)$ in \rf{dnew4} goes to zero for $\l \to \l_c$.
Phrased differently
the physical string tension will be infinite, except at special  critical
points $\l_c$ where the coefficient $\sg_0(\l_c)$ vanishes.
Assume that
\beq
\sg_0(\l) \sim (\l_c -\l)^{\a}.
\label{disc11}
\eeq
(The exponent $\a$ can be directly measured in our experiment by fitting
the behaviour of $\sg_0(\l)$ for $\l \to \l_c$ \rf{disc11}).
We do not expect a scaling of $d(\l)$ for $\l \to \l_c$.
 In the continuum limit we have $\m_R \to 0$ and
$\l \to \l_c$ with
\beq
\l_c - \l \sim \m_R^\rho.
\label{disc13}
\eeq
The critical exponents $\a$ and $\rho$ satisfy
\beq
\a\rho = 2\n ,
\label{disc14}
\eeq
In \rf{disc14} we assumed that both terms in \rf{dnew10}
contribute at the same order and we now see that the physical area
$A_{phys}$ in \rf{aux1} will be related to the ``bare'' area $A$ by
\beq
A_{phys} \sim \mu_R^{2\n} A
\label{disc9}
\eeq
and the physical string tension to the ``bare'' string tension as
\beq
\sg_{phys} \sim \frac{{\sg}(\l,\m)}{\m_R^{2\n } }
  = \frac{\sg_0(\l)}{\m_R^{2\n} } + d(\l)
\label{dnew10}
\eeq
showing explicitly that the physical string tension would be
infinite except at the critical points where $\sg_0(\l)$ vanishes.

The scaling assignments \rf{disc9} and \rf{dnew5} are in agreement
with the general scaling law relation of  mass critical exponents like
$\n$ to the Hausdorff dimension $d_H$ of the surfaces in target space:
\beq
\n = \frac{1}{d_H}.  \label{aux2}
\eeq
In fact one would define the Hausdorff dimension of the present ensemble
of surfaces by
\beq
N \sim A^{d_H/2}   \label{aux3}
\eeq
in the limit where $A_{phys}$ is kept fixed but the ``bare'' $A$ goes
to infinity. As we see $A_{phys}$ sets the scale for the divergence
of $A$ and $N$ for $\m_R \to 0$ (under the assumption that $\sg_0(\l)=0$):
\beq
A \sim \frac{A_{phys}}{\m_R^{2\n}},~~~~~
N \sim \frac{A_{phys}}{\m_R}, \label{aux4}
\eeq
which implies \rf{aux2}:
\beq
N \sim \left( A \right)^{1/(2\n)}. \label{bubu}
\eeq
Eq. \rf{bubu} is not in contradiction to \rf{dnew5}, but describes how the
limit $r \to 0$ should be taken in order to reach the continuum limit.

As we mentioned in the introduction an additional requirement for an
interesting scaling limit is the correct scaling of other physical
observables. An observable independent of the string tension is the
mass gap, defined in the grand canonical ensemble by the exponential
decay of the two-point function. In the standard approach one
fixes two points, separated by a distance
$y$, and sums over all surfaces passing through these points.
The transformation of the mass gap definition from the grand canonical
to the canonical ensemble was discussed in the preceeding section
(equations \rf{2.12a} to \rf{2.14a}) and we have the relation
\beq
N= \frac{\partial {\cal G}}{\partial \mu} =
\frac{\partial {m}(\l,\mu)}{\partial \mu} y.
\label{disc23}
\eeq

Let us now assume that the mass scales at the critical point $\mu_c (\l)$.
Opposite to the situation for the string tension we expect the bare mass
${m}(\l,\mu)$ to scale for $\m_R =\mu - \mu_c(\l) \to 0$
{\it for all} $\l$. Assume the
scaling is of the form
\beq
{m}(\l,\mu) \sim c(\l) \m_R^{\nu(\l)}
\label{disc24}
\eeq
Using \rf{disc24} one shows that $m(\l,N,y)$ defined by \rf{2.12a} to
\rf{2.14a}
has a scaling behaviour
\beq
m(\l,N,y) \sim D(\l)
           t^{\nu(\l)/(1-\nu(\l))},
\label{disc25}
\eeq
where $t = y/N$ is a scaling variable for the mass gap (analogous the $r$
variable for the string tension).
This formula allows us to determine the critical exponent $\nu(\l)$ from
a canonical simulation.

\section{The twisted boundary conditions}

In numerical simulations the systems necessarily have finite dimensions and all
measurements are subject to finite-size effects. In the ideal situation one
would like to use these effects to gain  additional information about the
critical properties of the system (like a finite-size scaling analysis). For
the
string tension measurements one may however meet effects of this kind, which
are
very difficult to estimate and which can strongly bias the measured critical
behaviour.
Following the discussion of the previous section, measurements of the string
tension would require introducing a boundary to the system. For a finite system
this may present a serious problem: the number of vertices belonging to the
boundary may be, and in practice is, a sizeable fraction of all vertices.
One has
also to provide some method of assigning vertices to the boundary. No obvious
solution seems to be at hand and any solution chosen may in fact influence the
results.

In this section we propose a method of avoiding these problems with the help of
twisted boundary conditions imposed on a surface with the topology of a torus.
In this case the parameter space of the surface with $N$ vertices
can be visualized as a plane, periodic in two non-parallel directions.
In each {\it elementary cell}
we have $N$ vertices, each vertex having infinitely many
periodic copies. The {\it elementary cells} can be numbered by two integers
$k_1$ and $k_2$.
Links can connect points both inside a single {\it cell}
and in the neighbouring {\it cells}. As before we assign
to each link $<ij>$ of a lattice in the parameter space
a vector $X_{ij}^{\m},\m = 1...D,$ in the target
space. We assume the map to be periodic. On a torus the embedding may be
nontrivial in the following sense. As a consequence of \rf{zero}
a sum of vectors along a closed path can take
values
\beq
E^{\m}(n_1,n_2) = n_1 E_1^{\m} + n_2 E_2^{\m}
\label{a1}
\eeq
with two constant vectors $E_1^{\m}$ and $E_2^{\m}$ and integers
$n_1$ and $n_2$ denoting the number of times the path winds around the two
axes of the torus. The fact that the vector $E^{\m}$ is non-zero does not
contradict the periodicity of $X_{ij}^{\m}$, because on a torus the closed
loops with non-zero values of $n_1$ and $n_2$ can not be contracted to a point.
We can therefore define non-trivial boundary conditions by choosing
two arbitrary vectors $E_1^{\m}$ and $E_2^{\m}$.
Such a choice of boundary conditions on a torus is the
analogue of the twisted boundary conditions used in the case of the gauge
theories, where the gauge fields are periodic up to non-trivial gauge
transformations.
The vectors $E_k^{\m}$ are topological invariants, they remain unchanged
if one or
more points along the path are shifted in the target space or if the
internal geometry of the surface is changed by a flip of some of its links.
The vector $E_k^{\m}$ does not depend on the point index $i$, but only on the
integer distance between the initial and final periods of the path. This is a
simple consequence of the periodicity of $X_{ij}^{\m}$.
For nonzero vectors $E_1^{\m}$ and $E_2^{\m}$, \rf{a1} means that
the coordinates $X_i^{\m}$ of the vertex $i$ are not strictly periodic but
depend also on the integer coordinates $k_1,k_2$ numbering the particular
{\it elementary cell} and are given by
\beq
X_i^{\m}(k_1,k_2)= X_i^{\m} + k_1 E_1^{\m} + k_2 E_2^{\m}.
\label{point}
\eeq
This again does not contradict the periodicity of the dynamical quantities,
which are functions of $X_{ij}^{\m}$.

One can check that this choice of boundary conditions corresponds to spanning
of
the system on a frame $E_1 \times E_2$. This can most easily be seen by
considering a minimum action configuration of the model. In the embedding
space all triangles lie in the two-plane spanned by the vectors $E_1$ and $E_2$
with the lengths of the
links minimizing the Gaussian part of the action. Due to
\rf{point} these lengths are non-zero for a non-zero frame. The advantage, as
compared to the situation where some points are assigned to the boundary is
that
now the boundary is not present, it becomes a translationally invariant concept
and the distribution of points is purely dynamical.

In our computation we have chosen the
vectors $E_1$ and $E_2$ to be perpendicular
with the lengths $L_1$ and $L_2$. For the string tension measurements we used
the square frame $L_1 = L_2$. For large $L_1$ and $L_2$ we expect the leading
behaviour to be on $A = L_1 L_2$. Translational invariance suggests that
subleading terms ($\propto L_1 + L_2$) are absent and we should observe only
small finite-size corrections of the form $L_1/L_2$. Another possible choice
can
be $L_1 \ne 0$ and large and $L_2 = 0$. This clearly corresponds to a
completely
different physical situation, where we  measure the analogue of the
point-point correlation function (or rather loop-loop correlation function)
with
the two loops kept at a distance $L_1$.

The implementation of the twisted boundary conditions
in the numerical simulations can present some practical
problems. In accordance with the discussion
above, one could be tempted to store
the lattice configurations,
using the link vectors $X_{ij}^{\m}$ rather than
the vertex coordinates $X_i^{\m}$. Such a parametrization is however
dangerous since it allows rounding errors to accumulate.
In particular, this can lead to a numerical violation of the
boundary conditions initially imposed, since both the vertex
coordinates and the triangulations are changed many times. To avoid this
problem we decomposed, following \rf{point}, the vector $X_{ij}^{\m}$ as
\beq
X_{ij}^{\m} = X_j^{\m} - X_i^{\m} + E_{ij}^{\m},
\label{num1}
\eeq
where $E_{ij}^{\m} = n_{ij}^1 E_1^{\m} + n_{ij}^2 E_2^{\m}$ and the integers
$n_{ij}^k$ are non-zero when the link $<ij>$ connects points in different
{\it elementary cells}. To store the configuration we need the
positions $X_j^{\m}$
of all points and two integers $n_{ij}^1,n_{ij}^2$ for every link.
Vectors $E_{ij}$ are additive, {\it i.e.}
$E_{ij} + E_{jk} + E_{ki} = 0$ for every triangle $<ijk>$. This implies that
also the $n_{ij}$ are additive.
The decomposition \rf{num1} is however not unique. For
arbitrary integers $l_i^1$ and $l_i^2$ one can perform a transformation
\bea
X_i^{\m} &\to& X_i^{\m} + l_i^1 E_1^{\m} + l_i^2 E_2^{\m},\\ \no
E_{ij}^{\m} &\to& E_{ij}^{\m}+ l_i^1 E_1^{\m} + l_i^2 E_2^{\m}
                             - l_j^1 E_1^{\m} - l_j^2 E_2^{\m},
\label{num2}
\eea
which leaves \rf{num1} invariant.
This transformation can be used to keep the $E_{ij}$ vectors bounded.

In the last section we defined the mass gap by the exponential decay
of the two-point function, the marked points on the surface separated
by a distance $y$ in target space.
Here we shall consider a different definition, making again use of the
twisted boundary conditions. This time we consider a system with a zero
projected area $A_p$, with only one non-zero vector $E_i$. We take
$E_1 = (y,0,...), E_2 = 0$. These boundary conditions permit to measure
a two-loop function, where two identical
loops are separated in a translationally invariant way by a distance $y$ and
where we sum over the loop size. The two-loop function should have the
same scaling behaviour as the two point function.

\section{Numerical results}

Let us now turn to the numerical simulations of the system.
For the measurements of the string tension and mass gap we will use
the twisted boundary conditions described it the last section. For
measurements of the specific heat where no extended frame in target  space
is needed these reduce to ordinary periodic boundary conditions
of surfaces with toroidal topology.
We are going to use essentially standard Monte Carlo techniques, but before
presenting the results, let us collect various computational aspects.

The degrees of freedom in the model are the vertex coordinates and
those describing the connectivity of the triangulation.
Both kinds we update by using
a local Metropolis algorithm. For the triangulation
we use the familiar flip procedure~\cite{kkm},
which is known to
be ergodic in the class of all triangulations
with the same topology. We use the standard restriction that
we do not allow non-trivial  loops of order one and two. In the
graph dual to a given triangulation, i.e. a $\phi^3$ graph, this restriction
corresponds to considering graphs without tadpoles and self-energy parts.
New values for the coordinates are proposed
by performing a heatbath or an $\omega=2$ overrelaxation
step~\cite{adler} with respect to the gaussian part of the action.
A Metropolis accept/reject step then takes
the curvature term into account.
We found this coordinate update
slightly more efficient than the one
in which the change in coordinates is generated
from a fixed distribution.
The gain we measured in decorrelation time
was somewhat less than a factor two.
The flip rate was close to 33\% in all our calculations.
The acceptance rate for the coordinate update
varied with the simulation parameters and was typically
similar to the flip rate. Its lowest value was 20\%,
at $\l=1.5$ with periodic boundary conditions.

The dynamical nature of the triangulation restricts the
possibilities to vectorize the simulation program.
We have therefore developed a program in which vectorization
is achieved by simulating several systems in parallel.
Our results have been obtained by simulating 64 parallel
systems on a CRAY-YMP. The
CPU time required per sweep per bond was 5.7 $\mu s$. For more
details on the algorithm and the program,
see ref.~\cite{ourpaper,program}.

While the simultaneous simulation of several systems
significantly reduces the CPU time needed per sweep,
it does, of course, not bring down the number of sweeps
needed to decorrelate configurations.
In fig.~1 we show estimates of the integrated
autocorrelation time, $\tau_{int}$, for the most non-local
quantity measured, the radius of gyration. $\tau_{int}$ refers
to the autocorrelation in one system, and the values given
are averages over the 64 systems considered.
We may note that
the increase in $\tau_{int}$ between
$N=144$ and 256, at fixed coupling, corresponds to
$\tau_{int}\propto N^{z'}$ with $z'$ roughly 1.6.
In our calculations, the number of sweeps used for
thermalization corresponds in most cases to 20$\tau_{int}$
and always to more than 10$\tau_{int}$. Measurements were typically
taken over 40$\tau_{int}$. This length of the runs is not very long
so one might worry about insufficient thermalization.
As a check, we therefore carried out a few additional much
longer runs. The longest one, at
$(N,\l)=(144,1.5)$ with periodic boundary conditions,
was two million sweeps,
corresponding to 700$\tau_{int}$. No significant change in the
observables was observed in these extended runs, and the statistical
errors scaled approximately as they should with increasing number of
iterations.
The integrated autocorrelation time did however
increase by 15\%.
The numbers given above for $\tau_{int}$ and its $N$ dependence
should therefore be used only as a rough guide.

In the analysis of the specific heat below we use the multi-histogram
technique by Ferrenberg and Swendsen~\cite{fs}.
This method allows continuation of results obtained
at one or more couplings to, in principle, arbitrary
couplings. With limited statistics this is, of course,
not true, and care is needed in selecting the
range for the continuation. We made the selection
in the following way, similar to that used
in ref.~\cite{bb}. We start by determining at each
simulated coupling two numbers $S_1$ and $S_2$ such
that the probabilities that $S_C<S_1$ and that $S_C>S_2$ are
both 25\%. We then perform a single-histogram continuation,
restricted to the coupling interval where the resulting value of $S_C$
lies between $S_1$ and $S_2$. With this restriction
we found that single-histogram results for the specific heat
agreed whenever the couplings overlapped. Final numbers were
obtained by combining the histograms according to the
method in ref.~\cite{fs}. The continuation
was carried out over a coupling range that could be covered
by the single-histogram intervals defined earlier.
The observed consistency between different single-histogram
results we take as another indication that
thermalization effects are under control.

Finally, we mention that the statistical errors quoted below
are jackknife errors~\cite{jack}, obtained by taking the
results from the different systems as 64 independent
measurements.

\subsection{Specific heat}

As observed first by Catterall~\cite{Catterall1}, using
spherical topology, the specific heat has a maximum
at a finite value of the coupling, $\l\approx 1.5$.
We have performed a detailed study of the specific heat
for the case of toroidal topology, with periodic
boundary conditions.
To accurately determine the location and height
of the maximum we use the multi-histogram technique by
Ferrenberg and Swendsen~\cite{fs}, as described above.
Fig.~2 shows the specific heat as a function of $\l$
for system sizes up to $N=576$. The curves are the results
from multi-histogram analysis, and the points show the results
at simulated couplings. In agreement with the results from
earlier studies using spherical topology,
we find a maximum in the specific heat near $\l=1.5$, the height
of which increases with increasing $N$.

The dependence of the position of the maximum, $\l_c^N$,
on the system size is shown in fig.~3.
The three data points with the largest values of $N$
are well described
by the form $\l_c^N=\l_c-{\rm const}N^{-\alpha}$ with
$\alpha=1/2$. However, acceptable fits to this form
can be obtained for a fairly wide range of $\alpha$ values.
As a result of this uncertainty about the form of the finite-size
corrections, we cannot get a very precise estimate
of the $N\rightarrow\infty$ limit $\l_c$. Fits with
$\chi^2<1$ are obtained for $0.35<\alpha<1.2$.
The corresponding bounds on the critical coupling are
$1.47<\l_c<1.53$.

How the height of the maximum, $C^N_{max}$, varies with the
system size is shown in fig.~4. The data strongly
suggest that $C^N_{max}$ remains finite in the limit
$N\rightarrow\infty$. The increase in $C^N_{max}$ with
increasing $N$ becomes slower at large $N$.
The finite-size correction
to the $N\rightarrow\infty$ value appears to vanish
faster than $1/N$ at large $N$.

\subsection{Mass gap}

In section~2 we saw how one can define a mass gap
measurement $m(\l,N,L)$ in the canonical ensemble.
In the thermodynamic limit $N\rightarrow\infty$,
this measurement gives information about the
mass gap $m(\l,\m)$, as defined from the exponential
decay of the grand canonical partition function. We have
\beq
m_{can}(\l,t)\equiv\lim_{N\rightarrow\infty,L/N=t} m(\l,N,L)=
m(\l,\m(t))\ ,
\label{mcan}\eeq
where $\m(t)$ is the solution to
\beq
{\partial m(\l,\m(t))\over \partial\m} ={1\over t}\ .
\label{mut}\eeq
Corresponding to the expected scaling $m(\l,\m)\sim
c(\l)\m_R^{\nu(\l)}$ at small $\m_R$, we have (eq.~\ref{disc25})
$m_{can}(\l,t)\sim D(\l)
t^{\nu(\l)/(1-\nu(\l))}$
at small $t\sim d(\l)\m_R^{1-\nu(\l)}$.
By verifying this scaling of $m_{can}(\l,t)$, we can compute
the critical exponent $\nu(\l)$.

The numerical calculations are restricted to finite $N$ and
it is therefore essential to keep finite-size effects under control
in the estimate of $m_{can}(\l,t)$. Let us give an estimate of
how large $N$ has to be taken at a given value of $t$.
The finite-size effects occur when
the asymptotic relation $\CG(\l,\m,L)\sim m(\l,\m)L$
is not fulfilled. This relation
is valid if terms of lower order in $L$ are negligible,
which should be the case when $m(\l,\m)L$ is large. We thus expect
finite-size corrections to be small if $m_{can}(\l,t)L$ is large.
This condition can in the scaling regime be written
as $t^{\nu(\l)/(1-\nu(\l))}L \gsim 1$, or
\beq
tN \gsim N^{\nu(\l)}.   \label{mbound}
\eeq

Let us briefly discuss the most likely finite-size effect to
the mass gap. The generic form of the two point function will be
(in the following we will suppress the dependence on the coupling
constant $\l$)
\beq
Z(\m,y) \sim y^{-\a} e^{-m(\m)y}   \label{xx1}
\eeq
This means that Gibbs free energy will be of the form
\beq
\CG(\m,y) = m(\m)y + \a \ln y   \label{xx2}
\eeq
and we would expect corrections of the form
\beq
m_{eff}(\m) \equiv \frac{\partial \CG}{\partial y} = m(\m) + \frac{\a}{y}.
\label{xx3}
\eeq
We can now translate this to the canonical ensemble and get
\beq
m_{can}(t,N)= m_{can}(t) + \frac{\a}{tN}.  \label{xx4}
\eeq

Our calculations were performed at two different couplings,
one in the crumpled phase, $\l=1.25$, and one close to
the maximum in the specific heat, $\l=1.5$. Fig.~5
shows the results for $m(\l,N,L)$, plotted against $t=L/N$.
In order to check the finite-size dependence, we carried out
simulations for different $N$ at some fixed values of $t$. As
expected from the discussion above, the finite-size
corrections turn out to be largest at small $t$.
The corrections are large for small $N$
($N=64$ data are omitted in the plot), but seem to
rapidly become smaller with increasing $N$.
The results for the largest values of $N$
agree well, which suggests that these can be taken as reasonable
approximations to the thermodynamic limit.

At large $t$ we find that
the results at the two couplings are similar,
which is expected due to the dominance of the gaussian term
in the action. The interesting region from the point of view of
scaling is, however, at small $t$,
and there we find a clear
coupling dependence.
The results at the three smallest values of $t$
are at both couplings consistent with scaling,
but with different exponents $\nu(\l)$.
Fitting the data points with largest $N$ to eq.~\rf{disc25} we get
$\nu(\l=1.25)=0.279(7)$ and $\nu(\l=1.5)=0.417(7)$, with $\chi^2$
near one in both cases. The errors are statistical only.
Clearly, one would like to confirm that these results reflect
the true scaling behaviour by going to
much smaller values of $t$.
Increasing finite-size effects and decorrelation times
have prevented us from doing so.
In the limited regime probed so far, we find
that the scaling assumption gives a good description
of fairly accurate data.

\subsection{String tension}

Our method to calculate the string tension is very similar
to that for the mass gap. Corresponding to eq.~\rf{mcan},
we have
\beq
\sg_{can}(\l,r)\equiv\lim_{N\rightarrow\infty,A/N=r} \sg(\l,N,A)=
\sg(\l,\m(r))\ ,
\label{scan}\eeq
where $\sg(\l,N,A)$ is the measurement defined in section 2 and
$\m(r)$ is given by
\beq
{\partial \sg(\l,\m(r))\over \partial\m} ={1\over r}\ .
\label{sr}\eeq
The major difference is in the expected scaling behaviour.
As explained in section 3, we expect
$\sg_{can}(\l,r)=\sg_0(\l) + \sg(\l)r^{2\nu/(1-2\nu)}$
with an in general non-zero $r$ independent term $\sg_0(\l)$.

In the same way as for the mass gap,
we can estimate for which parameter values one should expect
significant
finite-size effects.
They should be small if
$\sg_{can}(\l,r)A$ is large. If we assume the scaling form and
that $\sg_0(\l)=0$, then we can write this condition as
$r^{2\nu/(1-2\nu)}A \gsim 1$, or
\beq
rN \gsim N^{2\nu}.    \label{sbound}
\eeq
Repeating the arguments above we expect the finite-size scaling
corrections to be of the form
\beq
\sg_{can}(\l,r,N) = \sg_{can}(\l,r) + \frac{\a(\l)}{rN}. \label{yy1}
\eeq

We have carried out simulations at  three couplings
$\l=1.25, 1.4$ and 1.5, for system sizes up to $N=576$.
The results are summarized in fig.~6, where
$\sg(\l,N,A)$ is plotted against $r=A/N$. Finite-size
effects are large for $r\rightarrow 0$
and $\l\rightarrow \l_c$, as expected. Therefore, we have
not been able to measure the string tension
for very small $r$ at $\l=1.5$. It seems, however, that
$\sg_0(\l)$ could vanish near $\l=1.5$,
as required for an
interesting scaling behaviour. For this we need, in addition,
that the exponent in
\beq
\sg_{can}(\l,r)= \sg(\l) r^\omega
\label{omega}\eeq
satisfies $\omega=2\nu(\l)/(1-2\nu(\l))$, where $\nu(\l)$ is the
mass gap exponent. The value at $\l=1.5$ obtained above,
$\nu(\l=1.5)=0.417(7)$,
corresponds to $\omega=4.7\pm 0.5$.

Let us check if the measured string tension
at $\l=1.5$ shows this scaling behaviour.
To do so we first extrapolate the results to the thermodynamic
limit. In fig.~7 we show
$\sg(\l=1.5,A,N)$ at three fixed values of $r$ for $N=144,256,400$
and 576. For all three $r$,
we find that the  $1/N$ term
gives a statistically acceptable description of the finite-size
effects, in agreement with \rf{yy1}. The lines shown are fits to such a form.
We can now test for scaling by fitting the
extrapolated values to eq.~\rf{omega}.
We find that the data indeed are in good agreement with this
behaviour. This is illustrated in fig.~8.
The exponent obtained, $\omega=3.94\pm 0.06$,
is, moreover, close to the value
from the mass gap measurement.

This analysis suggests that there is a coupling close to $\l=1.5$
where the scaling form eq.~\rf{omega} is valid. It is however
likely that it is not exactly valid at $\l=1.5$. We therefore
want to check how the estimate of $\omega$ is affected
if we allow for a non-zero value of $\sg_0(\l=1.5)$. To this end
we performed a sequence of fits
for fixed values of $\omega$, having $\sg_0$ as a free
parameter. Fits with $\chi^2<1$ were obtained
for $3.8<\omega<4.5$.

The determination of $\omega$ also involves an assumption about
the precise form of the finite-size dependence. To get an idea
of the importance of this assumption we considered the
$N=576$ data directly, without any extrapolation procedure.
Repeating the type of fit
with $\omega$ fixed and $\sg_0$ as a free parameter,
we obtained this time $\chi^2<1$ for $3.4<\omega<4.0$.
We take this to suggest that the uncertainty in $\omega$ arising
from the $N\rightarrow\infty$ extrapolation is smaller than 0.5.
We then arrive at the estimate $3.4<\omega<5.0$, corresponding to
$0.38<\nu<0.42$.

The results presented above were all obtained in the crumpled phase or
close to the phase transition.
Unfortunately, we have no conclusive results for the flat phase.
For a flat surface the
minimum in the free energy $\CF$ is at a finite value of $r$.
The canonical string tension is therefore negative for sufficiently
small $r$. We tried to verify this
by performing some simulations
at $\l=1.75$. Due to large finite-size effects and decorrelation times
we could however not estimate
values for the limit of large $N$.

\subsection{Radius of gyration}

Another important quantity for the characterization
of the surfaces is the radius of gyration $R_g(\l,N)$,
which we, in the case of periodic boundary conditions, define as
\beq
R_g(\l,N) = {1\over N}\sum_i <\sqrt{(X_i^\mu)^2}>.
\label{rg}\eeq
The behaviour at large $N$,
\beq
R_g(\l,N)\sim N^{1/d_H(\l)},
\label{dh}\eeq
is taken to define the Hausdorff dimension $d_H(\l)$.
$d_H(\l)$ is said to be infinite if the growth with $N$ is
logarithmic.

The Hausdorff dimension is expected
to decrease with increasing $\l$, from
infinity at $\l=0$ to $d_H=2$ at large $\l$.
We here want to estimate the value at the phase transition,
$d_H(\l_c)$. To this end we consider the radius of
gyration at the maximum in the specific
heat, $R_g(\l_c^N,N)$, which we obtain by using again the
Ferrenberg-Swendsen method. The dependence of $R_g(\l_c^N,N)$
on $N$ is shown in the log-log plot fig.~9. If we assume that
$d_H(\l)$ varies smoothly across the transition, then
the slope at large $N$ should give us $1/d_H(\l_c)$.
Over the range of $N$ studied, we find that the
slope keeps decreasing with increasing $N$. The two
data points with largest $N$ correspond to a
value $d_H$=3.41(7).
We take this to suggest the lower bound $d_H(\l_c)>3.4$.

\section{Discussion}

We have shown that our data are compatible with a second order
transition at $\l=\l_c \approx 1.5$. This transition is characterized
by a vanishing string tension $\sg(\l)$ for $\l \to \l_c$
and a vanishing mass gap $m(\l)$ of the
two point function for $\l \to \l_c$. Furthermore
\beq
\frac{\sg(\l)}{m^2(\l)} \approx {\rm  const.}~~~~~{\rm for}~~~~ \l \to \l_c.
   \label{zz.1}
\eeq
This is precisely what we would expect in a string theory without tachyons.
It is not clear to us what is the effective continuum theory which
is compatible with such behaviour in three-dimensional target space.
As we explained in the introduction one part of the motivation for
investigating the critical behaviour of strings with extrinsic curvature
was the relation to fermionic strings where extrinsic curvature terms
arise in a natural way. From this point of view the results are
encouraging: Models with extrinsic curvature seem to exhibit non-trivial
critical behaviour. From the point of view of the superstring it would be
interesting to verify that this non-trivial behaviour persists for strings
embedded in higher dimensions. A further, most interesting problem to address
would be the inclusion of the Wess-Zumino-like terms mentioned in
the introduction. Such terms also arise naturally for fermionic strings
and could change the critical behaviour even further. It is unfortunately
not clear to us how these terms can be a part of a numerical study
like the one performed here since the Wess-Zumino term is complex.

In the context of conformal field theory the existence of a second order
transition gives rise to further interesting questions.
For the crystalline surfaces we have a well defined regular
two-dimensional lattice structure, and to
each lattice site $i$ we have associated $d$ real variables $x_i$  if
target space is $d$-dimensional.  The action between these variables is
{\it local}, although it is not polynomial due to the extrinsic curvature
term. In case there is a second order transition for a finite value  $\l_c$
of the extrinsic curvature coupling one would be tempted to conjecture
the existence of  an associated
conformal field theory, characterized by a central charge $c$.  If the
extrinsic curvature coupling is put to zero we just have a standard
gaussian field theory on the lattice and it has a gaussian
fixed point corresponding to $d$ free fields and a total central charge $d$.
In the present context the nature of this fixed point is a little unusual,
since it is an infrared fixed point and not the usual ultraviolet gaussian
fixed point  known for field theories in less than four dimensions.
In fact it seems as if the following effective action describes
the behaviour of the system for $\l < \l_c$ \cite{wheater}:
\beq
S_{eff}= \int d^2\xi \;x_\m(\xi)
\left[-m^2(\l)\partial^2(\xi) +\l \partial^4(\xi) \right] x_\m(\xi)  \label{7}
\eeq
where
\beq
m(\l) =m_0 (\l_c -\l).  \label{8}
\eeq
Eqs. \rf{7} and \rf{8} mean that the long range properties
of the system are determined by the gaussian part and for this reason one
would classify the first fixed point away from $\l=0$
as ultra-violet stable. If we naively apply Zamolodchikov's
$c$-theorem we can conclude that the central charge of the conformal
theory associated with ``crumpling'' transition has to be larger than
or equal $d$. In fact it is easy to calculate the central charge
at the fixed point, either analytically or numerically, using the
effective action \rf{7}-\rf{8}. One gets $c=2d$.
The numerical results for $d=3$ using the full action for
crystalline surfaces, not the approximation \rf{7},
indicate that the (effective)  central charge is
between zero and one \cite{kr}, in contradiction
with the $c$-theorem. Most likely we are
in fact considering a non-unitary conformal field theory in which case
the $c$-theorem is formally invalidated. However, one
expects a $c_{eff}$-theorem to apply even in that case and since the
central charge measured from the partition function in \cite{kr} {\it is}
the effective central charge $c_{eff}$ it does not change the
apparent contradiction. At least the situation calls for further
investigations and one should keep in mind that the action involving extrinsic
curvature is non-polynomial and it might be that we in this case are
mislead by the general principles, which do not necessarily apply.
Whatever the effective central charge is one can couple the lattice
system to two-dimensional quantum gravity and ask for the  properties
of the corresponding {\it random surface theory}. This is what we have
done, and since the situation already without coupling to gravity
(i.e. the crystalline surfaces) appears to be non-trivial, the same might
be the case when coupled to gravity. In fact the connection between
the conformal field theory on the world-sheet and the scaling properties
of the correlation functions in target space  has to be non-trivial if the
model without the Wess-Zumino term will describe a genuine surface
theory which is euclidean invariant and where the string tension scales.
Recent analytic results of Polchinski and Strominger indicate
that such non-trivial situations can indeed occur precisely when the
world-sheet variables $x_i$ contain non-polynomial terms \cite{ps}.
Whether our effective string theory  with a non-trivial scaling of
string tension and mass gap has any relation to the scenario suggested
in \cite{ps} is presently under investigation.

\vspace{36pt}
{\bf Acknowledgements}

The authors would like to thank the Hochleistungrechenzentrum in J\"{u}lich,
where most of our numerical work was performed. B.P. acknowledges several
inspiring discussions with Fran\c{c}ois David. J.J. thanks the Niels Bohr
Institute for a hospitality during his stay there and the DFG for the
financing of his visit to the Bielefeld University.

\newpage

\begin{center}

{\large \bf Figure Caption}

\end{center}

\vspace{24pt}

\begin{itemize}
\item[Fig.1]
Integrated autocorrelation time for the radius
of gyration against system size. The boundary conditions
are periodic, $y_1=y_2=0$.

\item[Fig.2]
The specific heat as a function of $\l$. The bands
give the results with errors of the multi-histogram analysis.

\item[Fig.3]
$\l_c^N$ against $N^{-1/2}$.

\item[Fig.4]
$C^N_{max}$ against $1/N$.

\item[Fig.5]The mass gap measurement $m(\l,N,L)$ plotted
against $t=L/N$. The straight lines are fits to eq.~\rf{disc25}.

\item[Fig.6]
The string tension measurement
$\sg(\l,N,A)$ plotted against $r=A/N$.
The lines connect points with the same $\l$ and largest $N$.

\item[Fig.7]
$\sg(\l,N,A)$ against $1/N$ at three fixed
values of $r$. The lines are fits to the form
$\sg(\l,N,A)=\sg_{can}(\l,r)+c(r)/N$.

\item[Fig.8]
$\sg(\l,N,A)$ against $r^\omega$ at $\l=1.5$.
The full circles are extrapolations to $N\rightarrow\infty$,
as explained in the text. For the exponent we have used the value
$\omega=3.94$, obtained
from a fit to eq.~\rf{omega}.

\item[Fig.9]
$R_g(\l_c^N,N)$ against $N$.

\end{itemize}

\end{document}